# Twisted MoSe$_2$ Homobilayer Behaving as a *Heterobilayer*

*Arka Karmakar[1*], Abdullah Al-Mahboob[2#], Natalia Zawadzka[1], Mateusz Raczyński[1], Weiguang Yang[3], Mehdi Arfaoui[4], Gayatri[1], Julia Kucharek[1], Jerzy T. Sadowski[2], Hyeon Suk Shin[3,5,6], Adam Babiński[1], Wojciech Pacuski[1], Tomasz Kazimierczuk[1], Maciej R Molas[1†]*

[1] Institute of Experimental Physics, Faculty of Physics, University of Warsaw, Pasteura 5, 02-093 Warsaw, Poland

[2] Center for Functional Nanomaterials, Brookhaven National Laboratory, Upton, NY 11973, USA

[3] Department of Chemistry, Ulsan National Institute of Science and Technology, Ulsan 44919, Republic of Korea

[4] Département de Physique, Faculté des Sciences de Tunis, Université Tunis El Manar, Campus Universitaire 1060 Tunis, Tunisia

[5] Center for 2D Quantum Heterostructures, Institute for Basic Science (IBS), Suwon 16419, Republic of Korea

[6] Department of Energy Science, Sungkyunkwan University, Suwon 16419, Republic of Korea

[*] arka.karmakar@fuw.edu.pl; karmakararka@gmail.com

[#] aalmahboo@bnl.gov

[†] maciej.molas@fuw.edu.pl

**Abstract:**

Heterostructures (HSs) formed by the transition-metal dichalcogenides (TMDCs) materials have shown great promise in next-generation optoelectronic and photonic applications. An artificially twisted HS, allows us to manipulate the optical, and electronic properties. With this work, we introduce the understanding of the complex energy transfer (ET) process governed by the dipolar interaction in a twisted molybdenum diselenide (MoSe$_2$) homobilayer *without* any charge-blocking interlayer. We fabricated an unconventional homobilayer (*i.e.*, HS) with a large twist angle by combining the chemical vapor deposition (CVD) and mechanical exfoliation (Exf.) techniques to fully exploit the lattice parameters mismatch and indirect/direct (CVD/Exf.) bandgap nature. This effectively weaken the charge transfer (CT) process and allows the ET process to take over the carrier recombination channels. We utilize a series of optical and electron spectroscopy techniques complementing by the density functional theory calculations, to describe a massive photoluminescence enhancement from the HS area due to an efficient ET process. Our results show that the electronically decoupled $MoSe_2$ homobilayer is coupled by the ET process, mimicking a 'true' heterobilayer nature.

Heterostructures (HSs) made by the vertical stacking of monolayers (1Ls) two-dimensional (2D) transition-metal dichalcogenides (TMDCs) have already been the trending topic[1] since the last decade due to their potential usage in the next generation ultrathin optoelectronic and photonic device applications.[2,3] Twisting the materials while forming the HS creates Moiré superlattices, which then modulate the electronic and optical properties.[4,5] Twisted 2D HSs have been extensively studied to understand the various Moiré physics; such as flat bands,[6] structure of Moiré exciton,[7] and inter/intralayer excitonic modulations,[8,9] just to name a few. Competing interlayer charge[10,11] (CT) and energy transfer[12–14] (ET) are the two main photocarriers relaxation routes in the HSs formed by the semiconducting TMDCs. Thus, understanding the competing interlayer processes and creating a comprehensive understanding is essential for the realization of TMDC-based optoelectronic applications. In simple words, CT happens when carriers tunnel from the higher-to-lower lying energy states. Whereas, the ET due to the dipole-dipole interaction, is analogically similar as the two pendulums with identical oscillating frequencies in classical mechanics. A few recent studies[15–18] have already explored the dependence of the twist angle on the interlayer CT process in 2D HSs. Nonetheless, up till now there has been no clear understanding of the complex ET process in a twisted TMDC HSs including the homobilayers. An earlier report[19] has shown an ET process in a homobilayer of tungsten disulfide ($WS_2$) separated by a thin hexagonal boron nitride (hBN) interlayer to block the indirect optical transition. Until now, an experimental study on the ET process in a twisted homobilayer *without* any charge-blocking interlayer, *i.e.*, in atomically close proximity, has yet to be reported. Since the traditional interlayer ET rate (~1 ps)[12] in TMDCs is found to happen at a much slower rate than the CT process (<50 fs).[10,18] Thus, CT is considered to be the dominating mechanism in the nanoscale regime (~1-2 nm range).

In this work, we explore an unusual ET coupling in a twisted molybdenum diselenide ($MoSe_2$) homobilayer made by the combination of chemical vapor deposition (CVD) and mechanical exfoliation (hereafter Exf.) technique. Instead of choosing the conventional HS fabrication routes, such as direct CVD growth[20,21] or polymer-based dry transfer methods,[22] we chose an unorthodox method of combining the CVD and Exf. technique. Such mixed approach allows us to exploit the different lattice parameters between the two materials produced by different growth techniques, create a homobilayer (*i.e.*, HS) with a clean large-area interface, and take the advantages of the direct/indirect (Exf./CVD) bandgaps overlap. Until now, this type of homobilayer with a twist angle has never been studied. We intentionally opt for the twisted HS with a large rotation (57°) combining with the lattice mismatch to minimize the charge tunneling[15,23] mediated by the excitonic wavefunctions overlap without any charge-blocking interlayer.[14] It is also important to mention that being an optically 'bright' material 1L $MoSe_2$ photoluminescence (PL) intensity significantly quenches at the room temperature as compared to the cryogenic temperature.[24] This study reveals a massive Exf. PL enhancement at room temperature from the HS area by a factor of ~7.6× due to the efficient interlayer ET process from the CVD to Exf. layer. We employ a series of optical and electron spectroscopy techniques at low (5 K) and room temperature (300 K), such as optical second harmonic generation (SHG), differential reflection contrast (RC), photoluminescence excitation (PLE), circularly polarized PL, time resolved PL (TR-PL), reflection high-energy electron diffraction (RHEED) and µ-beam low energy electron diffraction (µLEED). The experimental results, complementing by the density functional theory (DFT) calculations help us to demonstrate that the homobilayer effectively behaves as a 'heterobilayer', *i.e.*, each layer works independently, resulting no PL quenching effect due to the evolution of indirect bandgap as per the conventional twisted homobilayer structure.[8,25]

We grow isolated 1L $MoSe_2$ triangles on the ultraflat sapphire substrate using the CVD method. We then deterministically transfer the top Exf. $MoSe_2$ layer to create a HS (see Experimental Details for the sample fabrication procedure). Figure 1a shows the optical micrograph, while inset shows the schematic illustration of the sample's cross-section. The optical SHG technique has been proven to be a non-destructive method to determine the stacking angle in the artificial twisted 2D HSs.[26] Our SHG experiment (see Experimental Details) shows a ~57° twist angle between the two layers (Figure 1b). Figure 1c shows a graphic illustration of the 57° rotational angle between the hexagonal Brillouin Zones (BZs) of two 1Ls.

RHEED is an ideal tool to compare the reciprocal lattice parameters in thin films.[27] From the RHEED measurements (more details in the Experimental section) we find the CVD $MoSe_2$ lattice parameter to be ~3.25 Å (Figure 1d). Due to the different substrate condition and lack of reference point, we do not consider the Exf. $MoSe_2$ RHEED pattern on the $SiO_2$/Si substrate (Figure S1) for our analysis. Rather, we perform a µLEED measurement (Figure 1e) to find out the true lattice constant of ~3.34 Å from the Exf. flake (more in the Experimental details). The CVD growth[28] on the sapphire substrate could cause this strain-induced lattice mismatch as compared to the relaxed (Exf.) top layer. Later, we show how this ~2.8% lattice mismatch combining with the 57° twist angle between the two layers led us to observe an unprecedented ET process in the HS area.

The absorbance (Abs.) spectra (Figure 1f) were calculated from the differential RC measurements (see Experimental Details) using the following equation:[29]

$$Abs. = \frac{1}{4} \times (n_s^2 - 1)(\frac{R_s - R_{sub}}{R_s + R_{sub}})$$

where, $n_s = 1.77$ is the refractive index of the underlying sapphire substrate. The Abs. spectra show an overall red-shift in the CVD peaks positions as compared to the Exf. peaks at both the temperature regime (5 K and 300 K). The HS data shows a combination of peak from both the layers. At 5 K, the slight red-shift in the Exf. A peak from the HS area as compared to the isolated Exf. A peak could be a result due to change in the dielectric environment.

PLE provides material's quasi-absorption spectra and is an ideal tool for studying the interlayer ET process.[12,14] The low temperature (5 K) µ-PLE measurements (see Experimental Details) show a well separated narrow neutral ($X_o$) and bound excitonic emission ($X^-$) from the individual Exf. and HS area, and the CVD PL emission shows a typical[30] pronounced emission centered ~1.55 eV (Figure 2a-2c). We also see a phonon crossover[31] with the PL emission at near resonant A excitation in the Exf. PLE map (indicated by the white arrows in Figure 2b). Upon increasing the temperature to 300 K, all the peaks merged to an expected broad emission feature (Figure 2d-2f). The excitation and emission ranges in all the PLE measurements were kept constant to compare any temperature-induced shifts or changes in the spectra. 5 K PL emission at an excitation of ~1.88 eV, matching with the Exf. $MoSe_2$ B excitonic level (along the horizontal solid line in Figure 2a-2c), shows a slight enhancement of the $X_o$ emission by a factor of ~1.2× in the HS area as compared to the individual Exf. area (Figure 2g). At low temperature, the $X^-$ emission from the HS area was quenched as compared to the Exf. spectrum (Figure 2g), which may be a result of the PL filtering effect[32] due to the transfer of the bound states from the Exf. to CVD layer. Moreover, the reduced bound-state emission proves a good coupling between the individual layers. The room temperature (300 K) PL emission at an excitation energy matching with the Exf. $MoSe_2$ B level (~1.8 eV), shows a massive increase of the HS PL emission by a factor of ~7.6× (Figure 2h). The PL enhancement factor is defined as the ratio of the maximum integrated PL intensity from the HS area to the isolated Exf. emission, while all the experimental parameters remain the same. The PLE plot, which is basically a change of the emission landscape as a function of the excitation energies (along the vertical dotted line in Figure 2b-2c), shows a similar slight enhancement of the HS emission at 5 K (Figure 2i) and an overall massive enhancement at 300 K (Figure 2j) as compared to the Exf. emission. We would like to emphasize the fact that in our PL measurements, we do not observe any indirect emission due to the emergence of an indirect bandgap at the $K$-$\Gamma$ transition as per the traditional TMDC homobilayer studies.[8,25]

To verify the data reproducibility, we prepare two more HSs with different large twist angles, and both show a pronounced HS PL emission at room temperature (see Supplementary Information Figures S3-S4). Furthermore, we perform a room temperature PL intensity map at a resonant excitation of Exf. $MoSe_2$ B level (Figure S5) to confirm the uniformity of the pronounced PL emission throughout the entire HS area. These results show that the reported PL enhancement is not related to a sample-specific condition but due to a very unusual coupling between the two layers, as discussed in the later sections. It is important to mention that we only consider the Exf. PL enhancement in the HS area is due to the ET process from the CVD layer, as the HS PL emission shape, linewidth, and position perfectly match with the individual Exf. area (Figure S6). Also, ET from the CVD to Exf. layer happens only to the $X_o$ level. Because, $X^-$ emission is related to the local effect/intrinsic doping in the sample. While, the ET is mainly associated with the coupling between the non-localized (free) excitons. Thus, our main focus is to study the enhanced $X_o$ HS PL emission due to the ET from the CVD to the Exf. layer.

Circularly polarized light creates exciton (e-h pair) in a single valley based on the optical selection rule. These photocarriers can promote the intervalley scattering between the $K^+$ and $K^-$ valleys *via* different scattering mechanisms, mainly governed by the Coulomb interactions[33] or phonon-mediated scattering.[34] Thus, studying the circularly polarized PL emission and measuring the degree of circular polarization (*P*) provide us useful information about the valley population.[34] P is determined by the following equation:

$$P = \frac{I(\sigma^+) - I(\sigma^-)}{I(\sigma^+) + I(\sigma^-)}$$

where, $I(\sigma^+/\sigma^-)$ are the circularly polarized PL intensities. Circularly polarized excitation matching with the material's B excitonic level, results in PL emission from the material's A excitonic level from the same valley, conserving the photocarriers' total spin-state (Figure 3a). Low temperature (5 K) circularly polarized PL emission at an excitation energy of 1.88 eV (a technical discussion is provided in the Experimental Details) shows a low-to-high change of the *P* value for the neutral ($X_o$) emission from the CVD (55%), Exf. (76%) and HS (92%) area, respectively (Figure 3b-3d). Upon increasing the temperature to 300 K, the CVD area sees an increase (89%) and the Exf. area experiences a reduction (50%) in *P* (Figure 3e-3f). Whereas, the HS area demonstrates an unchanged *P* value of 92% (Figure 3g). An almost constant PL polarization with the increasing temperature was observed earlier.[35] However, the large circular polarization value of 92% from the HS area is to our knowledge the highest number reported till date from a mono- or bilayer TMDC system at room temperature.[36–38] A high *P* value at room temperature is a necessary condition for the future valleytronics applications based on the optically initialized *K*-valley polarization.[35] The high *P* value from the HS area indicates that the initially injected polarization primarily contributes in the PL emission.

The PL lifetimes (i.e., neutral excitonic decay time) determined by the TR-PL spectra (more details in the Experimental Section and Supplementary Information) show an usual behavior from the CVD and Exf. area, but, a completely surprising characteristic from the HS area at 300 K (Table 1 and Figure 3h-3i). At 7 K, the Exf. decay time undergoes a slight reduction as compared to the CVD PL lifetime, and the HS PL lifetime fells in between the two layers. Both the CVD and Exf. decay times at 300 K show a higher value due to the increased phonon-assisted intervalley scattering.[23,39] However, a prolonged HS decay time at room temperature, even more than the individual layers, suggests a different mechanism as discussed in the next section. The bound-state $X^-$ emissions from all the three regions show a longer decay time at low temperature as compared to the $X_o$ emission (Figure S7). Since the ET process happens faster than the thermalization of exciton distribution, the excess carrier relaxation due to the ET will dominate only by the radiative recombination of the neutral (intrinsic) exciton. Thus, this bound-state prolonged emission will be suppressed, and as a result we see a quenching of the HS $X^-$ emission in the PLE measurement (Figure 2g).

*Table 1: TR-PL Decay time obtained from the curve fitting as described in the Experimental Details.*

|  | 7 K | 300 K | |
| --- | --- | --- | --- |
|  | Decay Time (ps) | $\tau_1$ (ps) | $\tau_2$ (ps) |
| CVD | 2.866 ± 0.107 | 7.258 ± 0.359 | 54.988 ± 1.915 |
| Exf. | 2.081 ± 0.025 | 5.388 ± 0.158 | 49.153 ± 2.101 |
| HS | 2.503 ± 0.032 | 12.944 ± 1.088 | 96.803 ± 5.005 |

The DFT calculations (details are given in the Supplementary Information) reveal an indirect gap for the CVD MoSe$_2$ with energetically-lowest *K-Q* transition and a direct bandgap for the Exf. MoSe$_2$ layer located at the *K* points (Figure 4a-4b). The indirect bandgap of the CVD MoSe$_2$ is expected due to the smaller lattice parameter (~3.25 Å) compared to the relaxed (Exf.) MoSe$_2$ layer (~3.34 Å), mimicking the similar scenario under a compressive strain.[40,41] By comparing the band structures with the calculated work function (WF), we see that the two layers form an 'almost' type-II HS (Figure 4c). Now, after knowing the band alignment, we take into consideration of all the possible scenarios to explain our experimental results step-by-step as following:

I. The red-shift in the CVD excitonic level as compared to the Exf. layer (Figure 1f) is due to the similar effect in increased excitonic binding energy upon applying a compressive strain.[41]

II. The large twist angle combining with the lattice mismatch effectively weakens the interlayer charge tunneling.[15,23]

III. As shown in Figure 4d, upon resonantly excite electrons at the B transition in the CVD conduction band (CB) *K* valley (step 1), the electrons immediately scatter to the *Q* valley (step 2). These electrons then scatter back to the *K* valley *via* e-phonon scattering (step 3) before transferring to the Exf. layer *via* dipolar coupling (step 4) and then radiatively recombine to the ground state (step 5) creating excess PL emissions from the HS area.

IV. At 5 K, the kinetic energy ($E_K = \frac{3}{2}K_BT$, $K_B$ is the Boltzmann constant) of the electron is negligible, <1 meV. This small electron $E_K$, combining with the low phonon population make $Q{\rightarrow}K$ scattering (step 3) less probable at low temperature. On the other hand, at 300 K, the higher electron $E_K$ ~40 meV, pairing with the increased optical-phonon scattering[23] make the $Q{\rightarrow}K$ scattering (step 3) much effective. Thus, allowing an efficient ET coupling at room temperature. This scheme explains both the slight HS PL enhancement at 5 K, but a massive PL growth at 300 K (Figure 2i-2j).

V. The CVD band structure also helps us to interpret the increase in the CVD *P* value at 300 K (89%) as compared to the 5 K (55%), which is due to the surge in the intravalley scattering ($K{\leftrightarrow}Q$).

VI. Another interesting point is that at 300 K, the prolonged HS TR-PL lifetime (Table 1) combining with the band alignment (Figure 4c) indicates a both-way excitonic exchange (possibly multiple time) *via* ET process before radiatively recombine at the Exf. layer. Moreover, the excitonic up-conversion due to the WF mismatch (~120 meV) between the layers can be more probable when the large momentum phonon population is high, *i.e.*, at room temperature.[42]

Also, it is worth mentioning that close to AB (also called 2*H*) stacking helped us to achieve very strong interlayer coupling[8] in the HS area. Isolating the short-range interlayer CT and Dexter-type ET due to the electron wavefunction overlap is beyond the scope of this present work.

In conclusion, we experimentally show that the electronically decoupled large twisted $MoSe_2$ *homobilayer* is effectively coupled *via* interlayer ET process mediated by e-phonon interactions, which imitates a *heterobilayer* structure. This ET results in a 7.6× PL enhancement from the HS area at room temperature. In such HS, the large twist angle, together with the 2.8% lattice mismatch, effectively weakens the interlayer CT process, allowing the ET to take over the control of the photocarrier relaxation pathways *without* any charge-blocking interlayer. The DFT calculations prove an 'almost' type-II band alignment. This then explains the atypical prolonged excitonic lifetime from the HS area obtained at room temperature. Hence, we show that the $MoSe_2$ homobilayer behaves like a 'heterobilayer'. We also expect that the other Mo-based TMDC homobilayers ($MoS_2$ and $MoTe_2$) will have a similar effect as reported in the present work.

Our previous[13,14] and present works already showed that whenever there is an energetically resonant overlap between the optically excited states, ET dominates over the interlayer CT processes. Considering the fast interlayer CT timescale (<50 fs),[10,18] we can expect a faster ET period. However, finding the real ET rate will require a separate ultrafast study. Nonradiative ET in TMDC HSs is less explored as compared to the CT process due to the difficulties in direct observation of the dipole-dipole coupling, even with the modern state-of-the-art spectroscopy/microscopy techniques. In the last few years, the rapid advancement of the time-resolved angle-resolved photoemission spectroscopy (TR-ARPES) technique has already enabled the direct observation of the rapid formation of the exciton,[43,44] measurement of the excitonic wavefunction,[45] structure/formation of the Moiré exciton[7,46] and interlayer CT process[47] in semiconducting TMDC materials. We also believe that in the upcoming years the time- and energy-resolution of the state-of-the-art TR-ARPES will be greatly enhanced to become an ideal tool to study (visualize) the ET processes in the newly emerging 2D HSs. Another interesting point in our experiment is the difference in the PL enhancement factor between the PLE and circularly-polarized PL measurements. This clearly indicates a role of the inter- and intravalley transitions in the ET process and a detailed study is required to uncover the e-phonon scattering mechanism. We strongly believe that this work will be a stepping stone in understanding the complex ET processes in twisted HSs, which is ideal for designing next-generation optoelectronic, electronic, and photonic applications.

**Experimental Details:**

**HS Fabrication**

The triangular 1L MoSe$_2$ flakes on the sapphire substrates were grown using the CVD method as described in ref.[28] For the top MoSe$_2$ layer, first we exfoliated MoSe2 flakes from the bulk crystal onto the Gel-Pak (PDMS) films. After confirming the 1L nature of the exfoliated MoSe$_2$ flakes by the PL measurements, we carefully stacked the 1Ls on top of the CVD flakes using a home-built transfer setup. The samples were then annealed at 100 °C for ~30 minutes under ~40 mbar (30 Torr) pressure to make a good physical contact between the layers. We obtained commercially available bulk MoSe$_2$ crystals for the exfoliation from the HQ Graphene.

**Characterization**

The optical SHG experiment was performed using a Coherent Mira 900 Ti:Sapphire 76 MHz oscillator tuned to 800 nm. The polarization of the incoming fundamental beam was rotated using a half-waveplate placed directly in front of the microscope objective. The resulting 400 nm SHG beam was detected using a Czerny-Turner spectrometer equipped with a CCD camera. The weak HSs SHG signals from the 1st and 3rd samples (Figure 1b and Figure S4b) is a clear signature of destructive interferences of SH fields. Hence, confirming the 57° and 47° twist angle, respectively.[26] Whereas, the strong HS SHG signal from the 2nd sample (Figure S3b) shows a constructive interference of SH fields and confirms the 31° twist angle. In all the SHG data, the corresponding lobes from the 1L regions are indicated with arrows to determine the twist angles.

RHEED equipment is a part of the Molecular Beam Epitaxy reactor made by the SVT Associates, inc. RHEED patterns were observed on the CVD MoSe$_2$ sample grown on the sapphire substrate. Whereas, for the exfoliated MoSe$_2$ RHEED pattern we cleaved the commercially available bulk MoSe$_2$ crystals directly on the SiO$_2$/Si substrate to increase the material coverage on the substrate for better RHEED pattern visualization. Background pressure was kept at ~$10^{-9}$ Torr. Before the measurement, samples were quickly heated in ultra-high vacuum chamber up to 600 °C to degas the surface adsorbents and then immediately dropped down to 200 °C. RHEED patterns were acquired with 8kV accelerating electron energy. The CVD lattice parameter was obtained ~3.25 Å by using the sapphire as a reference.[48]

Exf. MoSe$_2$ sample was further studied employing the low-energy electron microscopy/photoelectron emission microscopy (LEEM/PEEM) technique.[49] To determine the lattice parameter of 1L Exf. MoSe$_2$ using µLEED pattern, Exf. flake was transferred onto Au coated mica. LEED pattern was recorded from selected 1.5 µ area of the sample (µLEED) in LEEM. For calibrating the µLEED pattern, Ru(0001)Ox surface was prepared *in-situ* in LEEM immediately after taking the MoSe$_2$ µLEED data. MoSe$_2$ µLEED pattern was scaled using O(2×1,1×2)-Ru(0001) spots (Figure S2) with identical optical LEEM setting and alignment. Obtained in-plane lattice parameter of Exf. MoSe$_2$ is ~3.34 Å.

We performed the differential RC measurements using a tungsten halogen lamp focused by a Mitutoyo Plan Apo SL 50x (N.A. 0.42) objective and directed into a spectrometer. Samples were loaded in a continuous flow cryostat and cooled with liquid helium (LHe). The differential reflectance is defined by $(R_s-R_{sub})/(R_s+R_{sub})$, where $R_s$ is the reflected light intensity from the MoSe$_2$ flakes and $R_{sub}$ from the sapphire substrate.

The µ-PLE experiments were performed using a super-continuum light source coupled with a monochromator as an excitation source. Samples were cooled down using the same cryostat used in the RC measurements. The incident light was focused on the samples using a Mitutoyo Plan Apo SL 100x (N.A. 0.55) objective. The average power on the sample was kept only ~13 µW (laser spot diameter ~1 µm) to avoid any high power induced non-linear effects from the sample. For the circularly polarized PL measurement we used the same setup at a fixed excitation of ~1.88 eV photons. λ/4 waveplates were used in the excitation and detection paths to control the left and right circularly polarized light. The λ/4 waveplate at the detection side was rotated to control either the same excitation/detection ($\sigma^+/\sigma^+$) or opposite excitation/detection ($\sigma^+/\sigma^-$) configuration.

TR-PL measurements were performed using a S20 synchro-scan streak camera. In these experiments, the sample was excited with an femtosecond Optical Parametric Oscillator (OPO) with 660 nm light at 7 K and 690 nm at 300 K, to match with the Exf. MoSe$_2$ B level at a repetition rate of 76 MHz. The sample was cooled down with LHe inside a continuous flow cryostat. We kept the average power on the sample ~30-40 µW.

**Data availability:**

All the necessary data to conclude the results are presented in the manuscript and supplementary file.

**Code availability:**

The technical details of the theoretical simulations are available from the corresponding authors upon reasonable request.

**Supporting Information:**

RHEED pattern of Exf. MoSe$_2$, µLEED pattern of RuO$_2$, PLE measurements of 31° and 47° twist samples, PL intensity map, normalized PL emission comparison, TR-PL fitting equations and spectra of X$^-$ emission, and details of the DFT calculations.

**References:**

(1) Geim, A. K.; Grigorieva, I. V. Van Der Waals Heterostructures. *Nature* **2013**, *499* (7459), 419–425. https://doi.org/10.1038/nature12385.

(2) Gong, C.; Zhang, Y.; Chen, W.; Chu, J.; Lei, T.; Pu, J.; Dai, L.; Wu, C.; Cheng, Y.; Zhai, T.; Li, L.; Xiong, J. Electronic and Optoelectronic Applications Based on 2D Novel Anisotropic Transition Metal Dichalcogenides. *Adv. Sci.* **2017**, *4* (12), 1700231. https://doi.org/10.1002/advs.201700231.

(3) Thakar, K.; Lodha, S. Optoelectronic and Photonic Devices Based on Transition Metal Dichalcogenides. *Mater. Res. Express* **2020**, *7* (1), 014002. https://doi.org/10.1088/2053-1591/ab5c9c.

(4) van der Zande, A. M.; Kunstmann, J.; Chernikov, A.; Chenet, D. A.; You, Y.; Zhang, X.; Huang, P. Y.; Berkelbach, T. C.; Wang, L.; Zhang, F.; Hybertsen, M. S.; Muller, D. A.; Reichman, D. R.; Heinz, T. F.;

Hone, J. C. Tailoring the Electronic Structure in Bilayer Molybdenum Disulfide via Interlayer Twist. *Nano Lett.* **2014**, *14* (7), 3869–3875. https://doi.org/10.1021/nl501077m.

(5) Zhang, C.; Chuu, C.-P.; Ren, X.; Li, M.-Y.; Li, L.-J.; Jin, C.; Chou, M.-Y.; Shih, C.-K. Interlayer Couplings, Moiré Patterns, and 2D Electronic Superlattices in MoS2/WSe2 Hetero-Bilayers. *Sci. Adv.* *3* (1), e1601459. https://doi.org/10.1126/sciadv.1601459.

(6) Naik, M. H.; Jain, M. Ultraflatbands and Shear Solitons in Moiré Patterns of Twisted Bilayer Transition Metal Dichalcogenides. *Phys. Rev. Lett.* **2018**, *121* (26), 266401. https://doi.org/10.1103/PhysRevLett.121.266401.

(7) Karni, O.; Barré, E.; Pareek, V.; Georgaras, J. D.; Man, M. K. L.; Sahoo, C.; Bacon, D. R.; Zhu, X.; Ribeiro, H. B.; O'Beirne, A. L.; Hu, J.; Al-Mahboob, A.; Abdelrasoul, M. M. M.; Chan, N. S.; Karmakar, A.; Winchester, A. J.; Kim, B.; Watanabe, K.; Taniguchi, T.; Barmak, K.; Madéo, J.; da Jornada, F. H.; Heinz, T. F.; Dani, K. M. Structure of the Moiré Exciton Captured by Imaging Its Electron and Hole. *Nature* **2022**, *603* (7900), 247–252. https://doi.org/10.1038/s41586-021-04360-y.

(8) Liu, K.; Zhang, L.; Cao, T.; Jin, C.; Qiu, D.; Zhou, Q.; Zettl, A.; Yang, P.; Louie, S. G.; Wang, F. Evolution of Interlayer Coupling in Twisted Molybdenum Disulfide Bilayers. *Nat. Commun.* **2014**, *5* (1), 4966. https://doi.org/10.1038/ncomms5966.

(9) Wu, B.; Zheng, H.; Li, S.; Ding, J.; He, J.; Zeng, Y.; Chen, K.; Liu, Z.; Chen, S.; Pan, A.; Liu, Y. Evidence for Moiré Intralayer Excitons in Twisted WSe2/WSe2 Homobilayer Superlattices. *Light Sci. Appl.* **2022**, *11* (1), 166. https://doi.org/10.1038/s41377-022-00854-0.

(10) Hong, X.; Kim, J.; Shi, S.-F.; Zhang, Y.; Jin, C.; Sun, Y.; Tongay, S.; Wu, J.; Zhang, Y.; Wang, F. Ultrafast Charge Transfer in Atomically Thin MoS2/WS2 Heterostructures. *Nat. Nanotechnol.* **2014**, *9* (9), 682–686. https://doi.org/10.1038/nnano.2014.167.

(11) Ceballos, F., Bellus, M. Z., Chiu, H-Y. & Zhao, H. Ultrafast Charge Separation and Indirect Exciton Formation in a MoS2–MoSe2 van der Waals Heterostructure. *ACS Nano* **8**, 12717–12724 (2014).

(12) Kozawa, D.; Carvalho, A.; Verzhbitskiy, I.; Giustiniano, F.; Miyauchi, Y.; Mouri, S.; Castro Neto, A. H.; Matsuda, K.; Eda, G. Evidence for Fast Interlayer Energy Transfer in MoSe2/WS2 Heterostructures. *Nano Lett.* **2016**, *16* (7), 4087–4093. https://doi.org/10.1021/acs.nanolett.6b00801.

(13) Karmakar, A.; Al-Mahboob, A.; Petoukhoff, C. E.; Kravchyna, O.; Chan, N. S.; Taniguchi, T.; Watanabe, K.; Dani, K. M. Dominating Interlayer Resonant Energy Transfer in Type-II 2D Heterostructure. *ACS Nano* **2022**, *16* (3), 3861–3869. https://doi.org/10.1021/acsnano.1c08798.

(14) Karmakar, A.; Kazimierczuk, T.; Antoniazzi, I.; Raczyński, M.; Park, S.; Jang, H.; Taniguchi, T.; Watanabe, K.; Babiński, A.; Al-Mahboob, A.; Molas, M. R. Excitation-Dependent High-Lying Excitonic Exchange via Interlayer Energy Transfer from Lower-to-Higher Bandgap 2D Material. *Nano Lett.* **2023**, *23* (12), 5617–5624. https://doi.org/10.1021/acs.nanolett.3c01127.

(15) Luo, D.; Tang, J.; Shen, X.; Ji, F.; Yang, J.; Weathersby, S.; Kozina, M. E.; Chen, Z.; Xiao, J.; Ye, Y.; Cao, T.; Zhang, G.; Wang, X.; Lindenberg, A. M. Twist-Angle-Dependent Ultrafast Charge Transfer in MoS2-Graphene van Der Waals Heterostructures. *Nano Lett.* **2021**, *21* (19), 8051–8057. https://doi.org/10.1021/acs.nanolett.1c02356.

(16) Zhang, Y.; Liu, T.; Fu, L. Electronic Structures, Charge Transfer, and Charge Order in Twisted Transition Metal Dichalcogenide Bilayers. *Phys. Rev. B* **2021**, *103* (15), 155142. https://doi.org/10.1103/PhysRevB.103.155142.

(17) Zimmermann, J. E.; Axt, M.; Mooshammer, F.; Nagler, P.; Schüller, C.; Korn, T.; Höfer, U.; Mette, G. Ultrafast Charge-Transfer Dynamics in Twisted MoS2/WSe2 Heterostructures. *ACS Nano* **2021**, *15* (9), 14725–14731. https://doi.org/10.1021/acsnano.1c04549.

(18) Zhu, H.; Wang, J.; Gong, Z.; Kim, Y. D.; Hone, J.; Zhu, X.-Y. Interfacial Charge Transfer Circumventing Momentum Mismatch at Two-Dimensional van Der Waals Heterojunctions. *Nano Lett.* **2017**, *17* (6), 3591–3598. https://doi.org/10.1021/acs.nanolett.7b00748.

(19) Xu, W.; Kozawa, D.; Zhou, Y.; Wang, Y.; Sheng, Y.; Jiang, T.; Strano, M. S.; Warner, J. H. Controlling Photoluminescence Enhancement and Energy Transfer in WS2:hBN:WS2 Vertical Stacks by Precise Interlayer Distances. *Small* **2020**, *16* (3), 1905985. https://doi.org/10.1002/smll.201905985.

(20) Lu, F.; Karmakar, A.; Shahi, S.; Einarsson, E. Selective and Confined Growth of Transition Metal Dichalcogenides on Transferred Graphene. *RSC Adv.* **2017**, *7* (59), 37310–37314. https://doi.org/10.1039/C7RA07772F.

(21) Gong, Y.; Lin, J.; Wang, X.; Shi, G.; Lei, S.; Lin, Z.; Zou, X.; Ye, G.; Vajtai, R.; Yakobson, B. I.; Terrones, H.; Terrones, M.; Tay, B. K.; Lou, J.; Pantelides, S. T.; Liu, Z.; Zhou, W.; Ajayan, P. M. Vertical and In-Plane Heterostructures from WS2/MoS2 Monolayers. *Nat. Mater.* **2014**, *13* (12), 1135–1142. https://doi.org/10.1038/nmat4091.

(22) Castellanos-Gomez, A.; Buscema, M.; Molenaar, R.; Singh, V.; Janssen, L.; van der Zant, H. S. J.; Steele, G. A. Deterministic Transfer of Two-Dimensional Materials by All-Dry Viscoelastic Stamping. *2D Mater.* **2014**, *1* (1), 011002. https://doi.org/10.1088/2053-1583/1/1/011002.

(23) Gao, S., Zhou, J-J., Luo, Y. & Bernardi, M. First-principles electron-phonon interactions and electronic transport in large-angle twisted bilayer graphene. Cond-mat.mtrl-sci at https://doi.org/10.48550/arXiv.2402.19453 (2024).

(24) Arora, A.; Nogajewski, K.; Molas, M.; Koperski, M.; Potemski, M. Exciton Band Structure in Layered MoSe2: From a Monolayer to the Bulk Limit. *Nanoscale* **2015**, *7* (48), 20769–20775. https://doi.org/10.1039/C5NR06782K.

(25) Grzeszczyk, M.; Szpakowski, J.; Slobodeniuk, A. O.; Kazimierczuk, T.; Bhatnagar, M.; Taniguchi, T.; Watanabe, K.; Kossacki, P.; Potemski, M.; Babiński, A.; Molas, M. R. The Optical Response of Artificially Twisted MoS2 Bilayers. *Sci. Rep.* **2021**, *11* (1), 17037. https://doi.org/10.1038/s41598-021-95700-5.

(26) Hsu, W-T.; Zhao, Z-A.; Li, L-J.; Chen, C-H.; Chiu, M-H.; Chang, P-S.; Chou, Y-C.; Chang, W-H. Second Harmonic Generation from Artificially Stacked Transition Metal Dichalcogenide Twisted Bilayers. *ACS Nano* **2014**, *8* (3), 2951–2958. https://doi.org/10.1021/nn500228r.

(27) Tempel, A.; Schumann, B. Determination of Lattice Parameters at Thin Epitaxial Layers by RHEED. *Krist. Tech.* **1979**, *14* (5), 571–574. https://doi.org/10.1002/crat.19790140510.

(28) Alexeev, E. M.; Ruiz-Tijerina, D. A.; Danovich, M.; Hamer, M. J.; Terry, D. J.; Nayak, P. K.; Ahn, S.; Pak, S.; Lee, J.; Sohn, J. I.; Molas, M. R.; Koperski, M.; Watanabe, K.; Taniguchi, T.; Novoselov, K. S.; Gorbachev, R. V.; Shin, H. S.; Fal'ko, V. I.; Tartakovskii, A. I. Resonantly Hybridized Excitons in Moiré Superlattices in van Der Waals Heterostructures. *Nature* **2019**, *567* (7746), 81–86. https://doi.org/10.1038/s41586-019-0986-9.

(29) Mak, K. F.; Sfeir, M. Y.; Wu, Y.; Lui, C. H.; Misewich, J. A.; Heinz, T. F. Measurement of the Optical Conductivity of Graphene. *Phys. Rev. Lett.* **2008**, *101* (19), 196405. https://doi.org/10.1103/PhysRevLett.101.196405.

(30) Mohapatra, P. K.; Deb, S.; Singh, B. P.; Vasa, P.; Dhar, S. Strictly Monolayer Large Continuous MoS2 Films on Diverse Substrates and Their Luminescence Properties. *Appl. Phys. Lett.* **2016**, *108* (4), 042101. https://doi.org/10.1063/1.4940751.

(31) Zinkiewicz, M.; Grzeszczyk, M.; Kazimierczuk, T.; Bartos, M.; Nogajewski, K.; Pacuski, W.; Watanabe, K.; Taniguchi, T.; Wysmołek, A.; Kossacki, P.; Potemski, M.; Babiński, A.; Molas, M. R. Raman Scattering Excitation in Monolayers of Semiconducting Transition Metal Dichalcogenides. *Npj 2D Mater. Appl.* **2024**, *8* (1), 2. https://doi.org/10.1038/s41699-023-00438-5.

(32) Lorchat, E.; López, L. E. P.; Robert, C.; Lagarde, D.; Froehlicher, G.; Taniguchi, T.; Watanabe, K.; Marie, X.; Berciaud, S. Filtering the Photoluminescence Spectra of Atomically Thin Semiconductors with Graphene. *Nat. Nanotechnol.* **2020**, *15* (4), 283–288. https://doi.org/10.1038/s41565-020-0644-2.

(33) Yu, T.; Wu, M. W. Valley Depolarization Due to Intervalley and Intravalley Electron-Hole Exchange Interactions in Monolayer MoS2. *Phys. Rev. B* **2014**, *89* (20), 205303. https://doi.org/10.1103/PhysRevB.89.205303.

(34) Kioseoglou, G.; Hanbicki, A. T.; Currie, M.; Friedman, A. L.; Gunlycke, D.; Jonker, B. T. Valley Polarization and Intervalley Scattering in Monolayer MoS2. *Appl. Phys. Lett.* **2012**, *101* (22), 221907. https://doi.org/10.1063/1.4768299.

(35) Lagarde, D.; Bouet, L.; Marie, X.; Zhu, C. R.; Liu, B. L.; Amand, T.; Tan, P. H.; Urbaszek, B. Carrier and Polarization Dynamics in Monolayer MoS2. *Phys. Rev. Lett.* **2014**, *112* (4), 047401. https://doi.org/10.1103/PhysRevLett.112.047401.

(36) Zhu, B.; Zeng, H.; Dai, J.; Gong, Z.; Cui, X. Anomalously Robust Valley Polarization and Valley Coherence in Bilayer WS2. *Proc. Natl. Acad. Sci.* **2014**, *111* (32), 11606–11611. https://doi.org/10.1073/pnas.1406960111.

(37) Godiksen, R. H.; Wang, S.; Raziman, T. V.; Rivas, J. G.; Curto, A. G. Impact of Indirect Transitions on Valley Polarization in WS2 and WSe2. *Nanoscale* **2022**, *14* (47), 17761–17769. https://doi.org/10.1039/D2NR04800K.

(38) Nayak, P. K., Lin, F-C., Yeh, C-H., Huang, J-S. & Chiu, P-W. Robust room temperature valley polarization in monolayer and bilayer WS2. *Nanoscale* **8**, 6035–6042 (2016).

(39) Sallen, G.; Bouet, L.; Marie, X.; Wang, G.; Zhu, C. R.; Han, W. P.; Lu, Y.; Tan, P. H.; Amand, T.; Liu, B. L.; Urbaszek, B. Robust Optical Emission Polarization in MoS2 Monolayers through Selective Valley Excitation. *Phys. Rev. B* **2012**, *86* (8), 081301. https://doi.org/10.1103/PhysRevB.86.081301.

(40) Cheng, X.; Jiang, L.; Li, Y.; Zhang, H.; Hu, C.; Xie, S.; Liu, M.; Qi, Z. Using Strain to Alter the Energy Bands of the Monolayer MoSe2: A Systematic Study Covering Both Tensile and Compressive States. *Appl. Surf. Sci.* **2020**, *521*, 146398. https://doi.org/10.1016/j.apsusc.2020.146398.

(41) Defo, R. K.; Fang, S.; Shirodkar, S. N.; Tritsaris, G. A.; Dimoulas, A.; Kaxiras, E. Strain Dependence of Band Gaps and Exciton Energies in Pure and Mixed Transition-Metal Dichalcogenides. *Phys. Rev. B* **2016**, *94* (15), 155310. https://doi.org/10.1103/PhysRevB.94.155310.

(42) Wu, Y-C. *et al.* Up- and Down-Conversion between Intra- and Intervalley Excitons in Waveguide Coupled Monolayer WSe2. *ACS Nano* **14**, 10503–10509 (2020).

(43) Madéo, J.; Man, M. K. L.; Sahoo, C.; Campbell, M.; Pareek, V.; Wong, E. L.; Al-Mahboob, A.; Chan, N. S.; Karmakar, A.; Mariserla, B. M. K.; Li, X.; Heinz, T. F.; Cao, T.; Dani, K. M. Directly Visualizing the Momentum-Forbidden Dark Excitons and Their Dynamics in Atomically Thin Semiconductors. *Science* **2020**, *370* (6521), 1199–1204. https://doi.org/10.1126/science.aba1029.

(44) Dong, S.; Puppin, M.; Pincelli, T.; Beaulieu, S.; Christiansen, D.; Hübener, H.; Nicholson, C. W.; Xian, R. P.; Dendzik, M.; Deng, Y.; Windsor, Y. W.; Selig, M.; Malic, E.; Rubio, A.; Knorr, A.; Wolf, M.; Rettig, L.; Ernstorfer, R. Direct Measurement of Key Exciton Properties: Energy, Dynamics, and Spatial Distribution of the Wave Function. *Nat. Sci.* **2021**, *1* (1), e10010. https://doi.org/10.1002/ntls.10010.

(45) Man, M. K. L.; Madéo, J.; Sahoo, C.; Xie, K.; Campbell, M.; Pareek, V.; Karmakar, A.; Wong, E. L.; Al-Mahboob, A.; Chan, N. S.; Bacon, D. R.; Zhu, X.; Abdelrasoul, M. M. M.; Li, X.; Heinz, T. F.; da Jornada,

F. H.; Cao, T.; Dani, K. M. Experimental Measurement of the Intrinsic Excitonic Wave Function. *Sci. Adv.* 7 (17), eabg0192. https://doi.org/10.1126/sciadv.abg0192.

(46) Schmitt, D.; Bange, J. P.; Bennecke, W.; AlMutairi, A.; Meneghini, G.; Watanabe, K.; Taniguchi, T.; Steil, D.; Luke, D. R.; Weitz, R. T.; Steil, S.; Jansen, G. S. M.; Brem, S.; Malic, E.; Hofmann, S.; Reutzel, M.; Mathias, S. Formation of Moiré Interlayer Excitons in Space and Time. *Nature* **2022**, *608* (7923), 499–503. https://doi.org/10.1038/s41586-022-04977-7.

(47) Bange, J. P.; Schmitt, D.; Bennecke, W.; Meneghini, G.; AlMutairi, A.; Watanabe, K.; Taniguchi, T.; Steil, D.; Steil, S.; Weitz, R. T.; Jansen, G. S. M.; Hofmann, S.; Brem, S.; Malic, E.; Reutzel, M.; Mathias, S. Probing Electron-Hole Coulomb Correlations in the Exciton Landscape of a Twisted Semiconductor Heterostructure. *Sci. Adv. 10* (6), eadi1323. https://doi.org/10.1126/sciadv.adi1323.

(48) Kurlov, V. N. Sapphire: Properties, Growth, and Applications. In *Encyclopedia of Materials: Science and Technology*; Buschow, K. H. J., Cahn, R. W., Flemings, M. C., Ilschner, B., Kramer, E. J., Mahajan, S., Veyssière, P., Eds.; Elsevier: Oxford, 2001; pp 8259–8264. https://doi.org/10.1016/B0-08-043152-6/01478-9.

(49) E Bauer. Low Energy Electron Microscopy. *Rep. Prog. Phys.* **1994**, *57* (9), 895. https://doi.org/10.1088/0034-4885/57/9/002.

**Acknowledgements:**

This work has been supported by the National Science Centre, Poland (Grant No. 2022/47/D/ST3/02086). W.Y. and H.S.S. also acknowledge the support from the National Research Foundation of Korea (Grant No. NRF-2021R1A3B1077184) and Institute for Basic Science (IBS-R036-D1). This research used resources of the Center for Functional Nanomaterials which is U.S. Department of Energy (DOE) Office of Science facility at the Brookhaven National Laboratory, under Contract No. DE-SC0012704.

**Author contributions:**

A.K. conceived the project. A.K. and M.R.M. designed the experiments. W.Y. and H.S.S. synthesized the CVD flakes. A.K. fabricated the HSs. A.K. and N.Z. performed the RC and PLE experiments. M.R. and T.K. did the SHG and TR-PL measurements. J.K. and W.P. conducted the RHEED experiment. A.A.M. and J.T.S. carried out the µLEED investigation. A.K., A.A.M., M.R., G. and T.K. analyzed the data. A.A.M. and M.A. performed the theoretical calculations. A.K., A.A.M., M.R.M., J.T.S. and A.B. interpreted the results. A.K. wrote the manuscript with responses taken from all the co-authors.

**Competing interests:**

The authors declare no competing financial interests.

**Declaration:**

We did not use any AI software in the data analysis or writing any part of the manuscript.

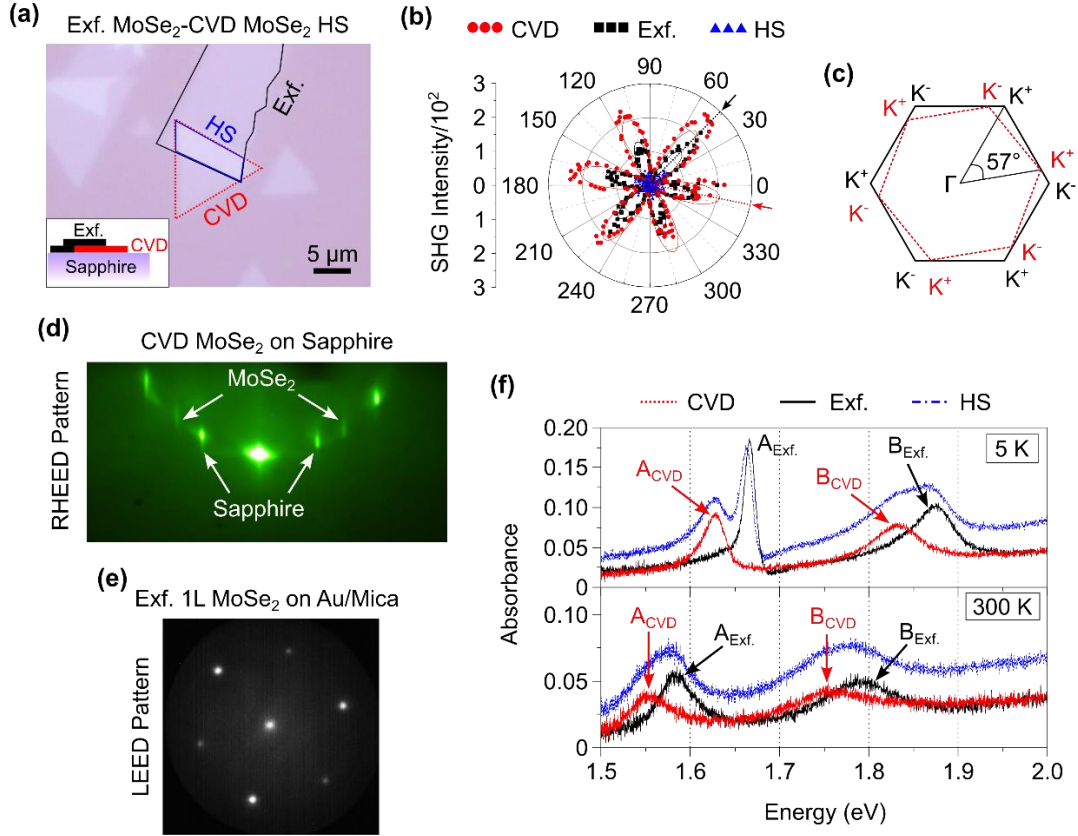

*Figure 1: Material characterization. (a) Optical image of the twisted monolayers (1Ls) Exfoliated (Exf.) MoSe$_2$-chemical vapor deposition (CVD) MoSe$_2$ heterostructure (HS) on the ultraflat sapphire substrate. Inset is the schematic illustration of the side-view of the sample. (b) Optical second harmonic generation (SHG) intensity polar plot from the sample to identify the rotational angle between the individual (1Ls). SHG intensities were divided by a factor of $10^2$. The angle difference between the two main lobes of the CVD and Exf. Layers are found to be ~57° (c) Schematic illustration of the 57° twisted Brillouin Zones (BZs) of the two layers as determined from the SHG measurements. (d)-(e) Reflection high-energy electron diffraction (RHEED) patterns from the CVD layer on sapphire and μLEED pattern of 1L Exf. Flake on the Au/Mica substrate taken with 40 eV electron beam, respectively. We find that there is a ~2.8% lattice parameters mismatch between the two materials. (f) Absorbance spectra converted from the differential reflection contrast (RC) measurements taken at cryogenic temperature (5 K) and room temperature (300 K), top and bottom panel, respectively. CVD excitonic peak positions show a constant red-shift as compared to the Exf. peak at both the temperature regime. HS peaks show summation of peaks from both the layers.*

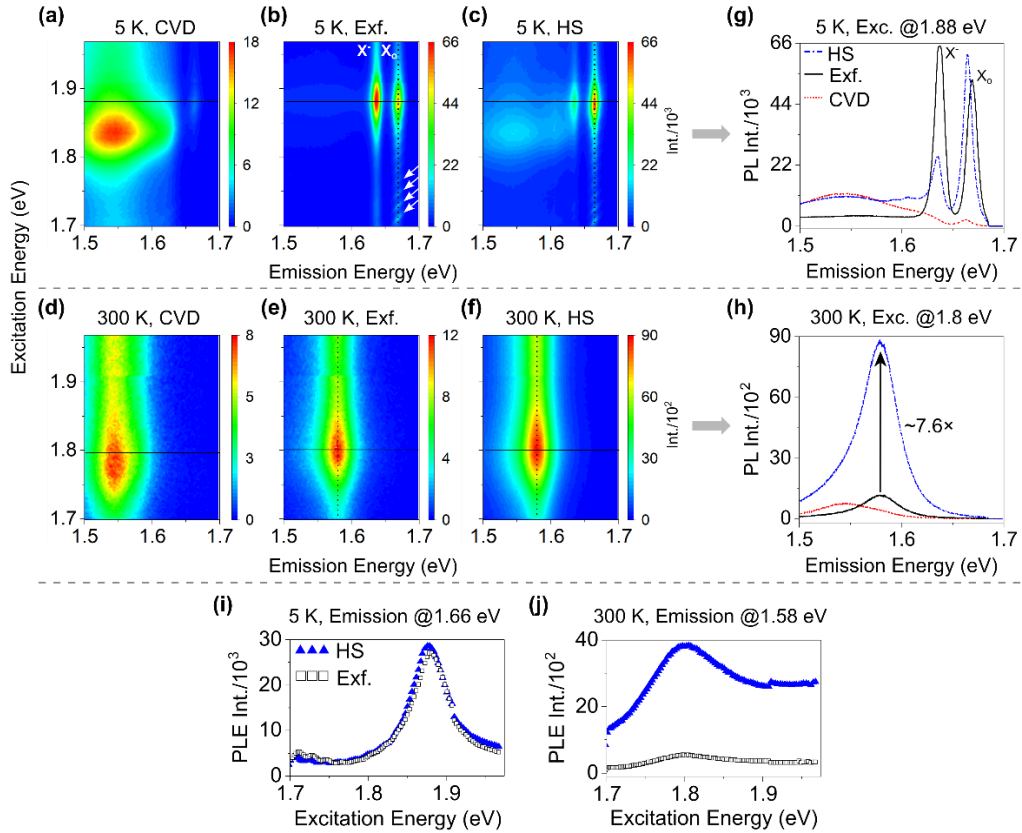

*Figure 2: PLE measurements. (a)-(c) Low temperature (5 K) photoluminescence excitation (PLE) measurements taken from the CVD, Exf. and HS area, respectively. Intensities were divided by a factor of $10^3$. In Figure (b) the white arrows indicate the phonon lines crossing the PL emission. (d)-(f) Room temperature (300 K) PLE measurements from the CVD, Exf. and HS area, respectively. Intensities were divided by a factor of $10^2$. (g) 5 K Photoluminescence (PL) emission plot at 1.88 eV excitation energy, i.e. along the horizontal solid line in (a)-(c). HS neutral exciton peak emission ($X_o$) shows an enhancement of ~1.2× as compared to the Exf. neutral exciton peak emission. (h) 300 K PL emission plot at 1.8 eV excitation energy, i.e. along the horizontal solid line in (d)-(f). HS neutral exciton peak emission shows an enhancement of ~7.6× as compared to the Exf. neutral exciton peak emission. (i) 5 K PLE plot from the Exf. and HS area at 1.66 eV emission energy (along the vertical dotted line in (b)-(c)). HS emission shows a slight PL increment as compared to the Exf. layer throughout the entire excitation range. (j) 300 K PLE plot from the Exf. and HS area at 1.58 eV emission energy (along the vertical dotted line in (e)-(f)). HS emission shows a massive PL increment as compared to the Exf. layer throughout the entire excitation range.*

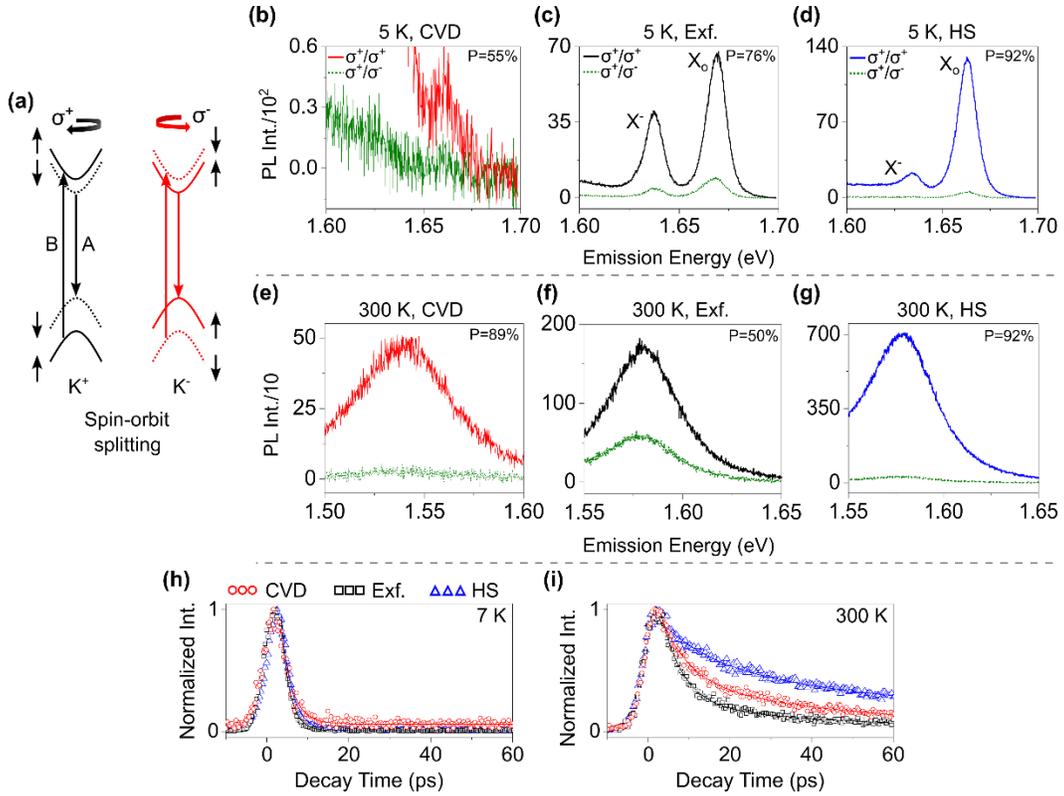

*Figure 3: Circularly polarized PL and TR-PL. (a) Schematic illustration of the circularly polarized PL emission at a resonant excitation of Exf. MoSe$_2$ B level. Valley-selective right ($\sigma^+$) and left ($\sigma^-$) circularly polarized excitations/emissions are marked with the black and red lines, respectively. (b)-(d) 5 K circularly polarized PL emission at an excitation of 1.88 eV from the CVD, Exf. and HS area, respectively. In each of the layer the PL comparisons were made with the same excitation/detection ($\sigma^+/\sigma^+$) and opposite excitation/detection ($\sigma^+/\sigma^-$) configuration to obtain the degree of circular polarization (P) of the neutral exciton ($X_o$). PL intensities were divided by a factor of $10^2$. (e)-(g) Similar circularly polarized PL emission at 300 k with 1.88 eV excitation of CVD, Exf. and HS region, respectively. PL intensities were divided by a factor of 10. (h)-(i) Time resolved PL (TR-PL) spectra from the three regions measured at 7 K and 300 K, respectively. At 7 K, the decay times from the CVD, Exf. and HS areas are 2.866 ps, 2.081 ps and 2.503 ps, respectively. At 300 K, the spectra were fitted using a bi-exponential decay function to obtain the fast ($\tau_1$) and slower ($\tau_2$) time constants. $\tau_1$ and $\tau_2$ values from the CVD, Exf. and HS areas are 7.258 ps, 54.988 ps; 5.388 ps, 49.153 ps; and 12.944 ps, 96.803 ps, respectively. The hollow symbols represent the real signal and the solid lines are the fitted data.*

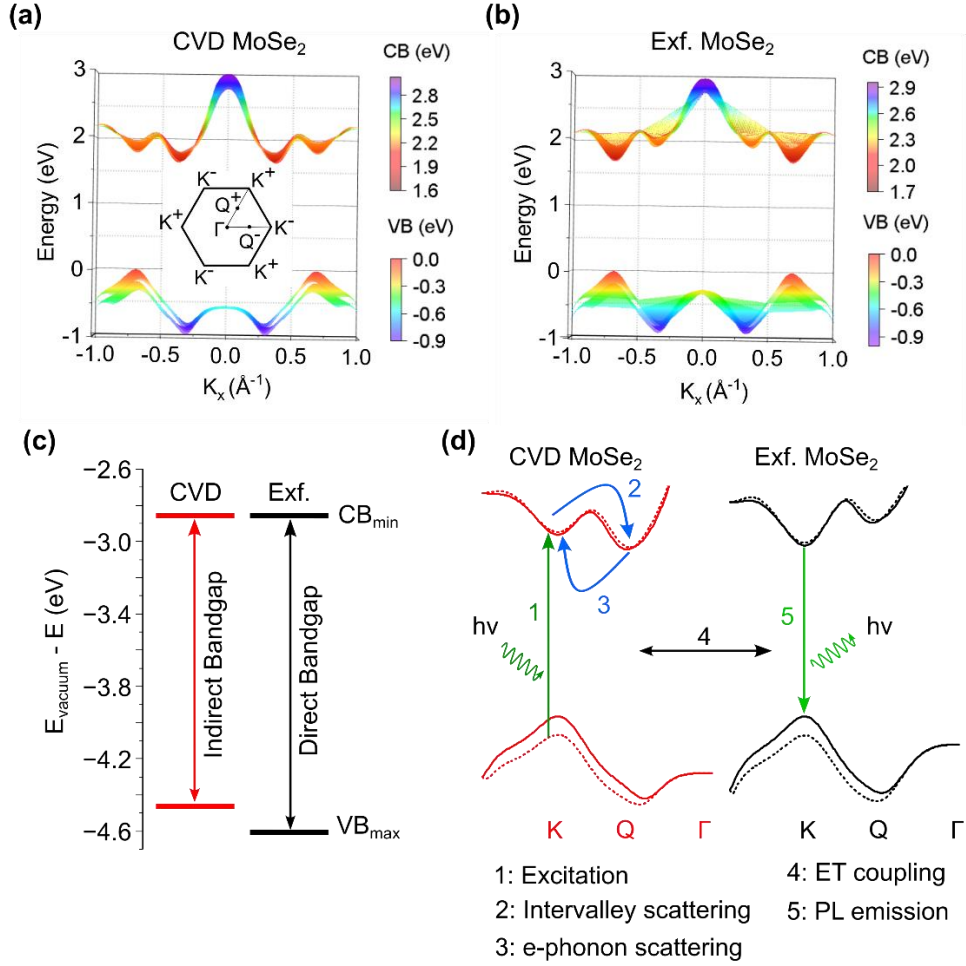

*Figure 4: **Band alignment and photocarrier relaxation pathways.** (a)-(b) Density functional theory (DFT) computed 3D energy band diagram of CVD and Exf. layer, respectively. The DFT computed energy bandgaps were matched with the energy levels obtained from the absorbance spectra. The CVD layer shows an indirect bandgap at K-Q transition. Whereas, the Exf. layer shows a direct bandgap at K-K transition. Inset of (a) shows the cut in the BZ along which the computed band structures were plotted. (c) Schematic illustration of the band alignment between the two layers matching with the calculated work function (WF) indicates an 'almost' type-II HS configuration. (d) Graphical representation of the photocarrier relaxation pathways in the HS at resonant B excitation. The spin-orbit splitting band structures are represented by the solid and dotted lines. For a simplified view, step-by-step transitions are labeled in the diagram.*

# SUPPORTING INFORMATION: Twisted MoSe$_2$ Homobilayer Behaving as a *Heterobilayer*

*Arka Karmakar[1*], Abdullah Al-Mahboob[2#], Natalia Zawadzka[1], Mateusz Raczyński[1], Weiguang Yang[3], Mehdi Arfaoui[4], Gayatri[1], Julia Kucharek[1], Jerzy T. Sadowski[2], Hyeon Suk Shin[3,5,6], Adam Babiński[1], Wojciech Pacuski[1], Tomasz Kazimierczuk[1], Maciej R Molas[1†]*

[1] Institute of Experimental Physics, Faculty of Physics, University of Warsaw, Pasteura 5, 02-093 Warsaw, Poland

[2] Center for Functional Nanomaterials, Brookhaven National Laboratory, Upton, NY 11973, USA

[3] Department of Chemistry, Ulsan National Institute of Science and Technology, Ulsan 44919, Republic of Korea

[4] Département de Physique, Faculté des Sciences de Tunis, Université Tunis El Manar, Campus Universitaire 1060 Tunis, Tunisia

[5] Center for 2D Quantum Heterostructures, Institute for Basic Science (IBS), Suwon 16419, Republic of Korea

[6] Department of Energy Science, Sungkyunkwan University, Suwon 16419, Republic of Korea

[*] arka.karmakar@fuw.edu.pl; karmakararka@gmail.com

[#] aalmahboo@bnl.gov

[†] maciej.molas@fuw.edu.pl

**RHEED pattern of Exfoliated (Exf.) MoSe$_2$:**

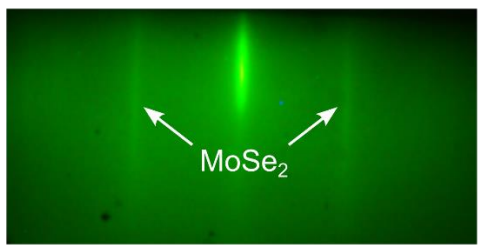

*Figure S5: RHEED pattern of the Exf. MoSe$_2$ flakes on SiO$_2$/Si substrate. Despite the high crystal quality of exfoliated flakes, the observed diffraction lines are broader than in the case of epitaxial CVD flakes (Fig. 1d in the main manuscript), due to the random orientation of the Exf. flakes.*

**µLEED pattern of O(2x1, 1x2)-Ru(0001):**

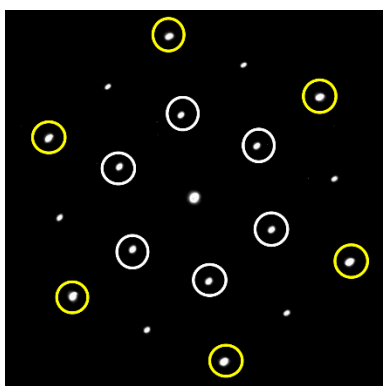

*Figure S6: µLEED pattern from the ruthenium-oxide surface taken with 40 eV electron beam. O(2×1,1×2)-Ru(0001) formed by oxidation of Ru(0001) surface. Yellow circles represent the Ru(**1×1**) spots and white circles represent the O(2×1,1×2) spots.. This surface was used to calibrate the reciprocal space to obtain the Exf. MoSe$_2$ lattice parameter.*

**PLE measurements on the 2$^{nd}$ HS with ~31° twist angle:**

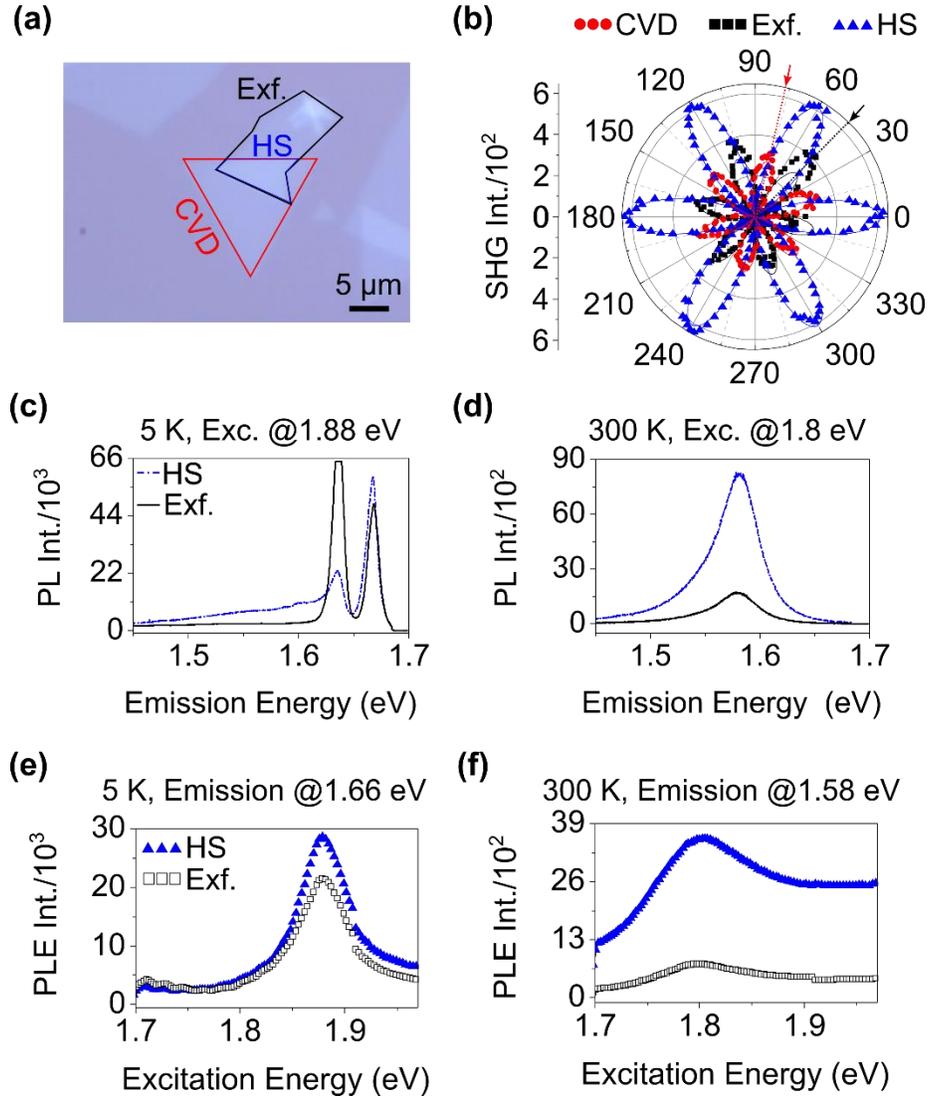

*Figure S3: (a) Optical micrograph of the 2$^{nd}$ MoSe$_2$ homobilayer fabricated by the Exf. and CV technique. (b) Optical SHG measurements show ~31° rotation between the two layers. (c)-(d) PL spectra at an excitation matching with the Exf. MoSe$_2$ B level at 5 K and 300 K, respectively. At low temperature the neutral PL emission from the HS area shows a slight enhancement as compared to the Exf. layer. Whereas, at room temperature HS PL shows ~4.8× intensity enhancement. (e)-(f) PLE plots show a similar increase in the PL emission throughout the entire excitation range at 5 K and 300 K, respectively.*

**PLE measurements on the 3rd HS with ~47° twist angle:**

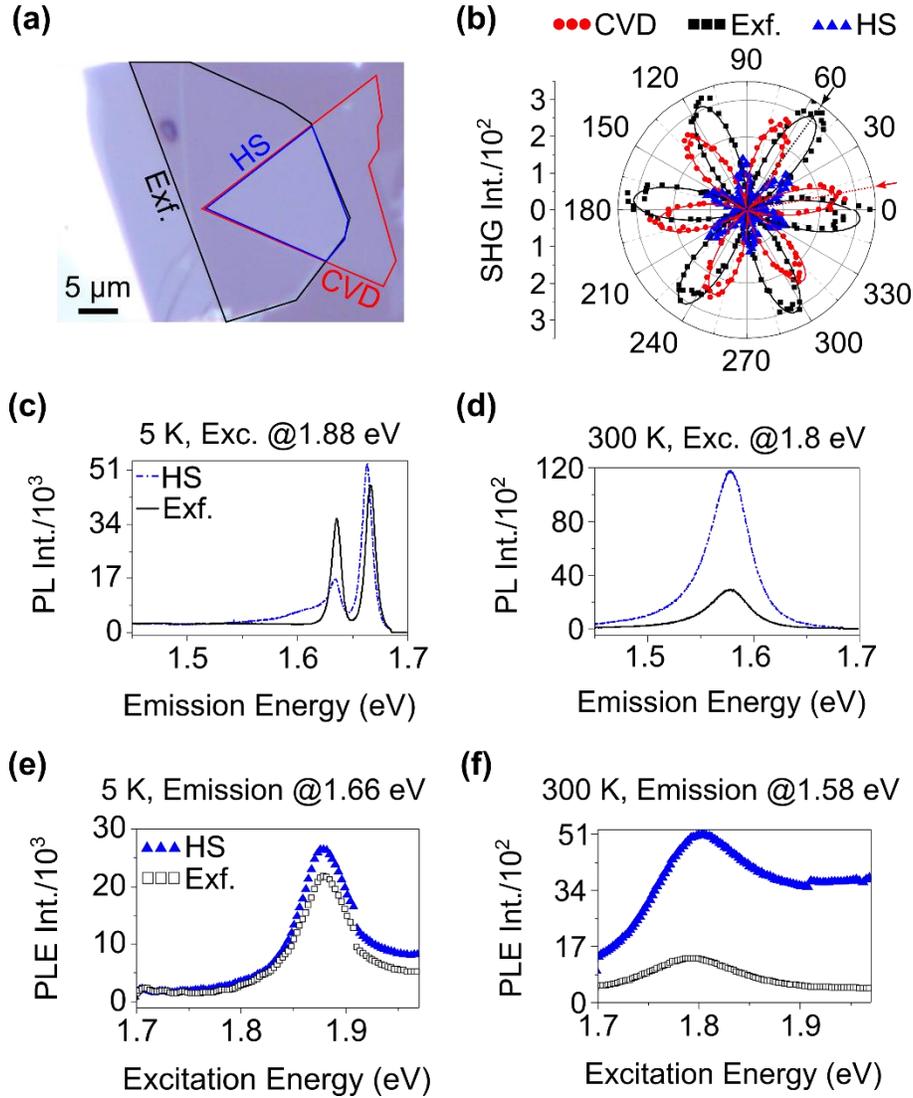

*Figure S4: (a) Optical micrograph of the 3rd MoSe$_2$ homobilayer fabricated by the Exf. and CV technique. (b) Optical SHG measurements show ~47° rotation between the two layers. (c)-(d) PL spectra at an excitation matching with the Exf. MoSe$_2$ B level at 5 K and 300 K, respectively. At low temperature the neutral PL emission from the HS area shows a slight enhancement as compared to the Exf. layer. Whereas, at room temperature HS PL shows ~4× intensity enhancement. (e)-(f) PLE plots show a similar increase in the PL emission throughout the entire excitation range at 5 K and 300 K, respectively.*

**PL intensity map of the 3rd HS with an excitation of 1.8 eV:**

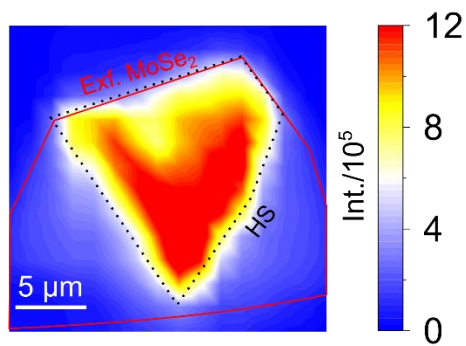

*Figure S5: Room temperature PL map of the total integrated intensity from the 3rd sample clearly shows an overall massive enhancement throughout the entire HS area. We note that, there is a slight non-uniformity in the homogeneous intensity distribution, which is typical for an exfoliated flake. However, the PL intensity from the HS is still higher as compared to the surrounding isolated layers.*

**Normalized PL emission comparison:**

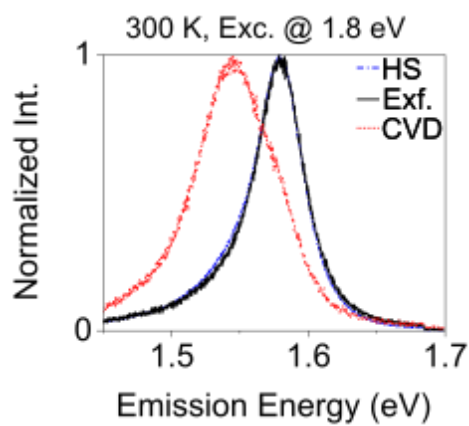

*Figure S6: Normalized PL emission comparison under 1.8 eV excitation at room temperature. HS PL peak position, linewidth and shape perfectly matches with the Exf. PL emission.*

**Fitting equations of the TR-PL spectra:**

Low temperature (7 K) spectra were fitted using a mono-exponential decay function:

$$y = y_o + A_1 \exp\left(-\frac{x - x_o}{\tau_1}\right)$$

where, $y_o$ is the y offset, $x_o$ is the x offset, $A_1$ is the amplitude and $\tau_1$ is the time constant. Whereas, the room temperature (300 K) spectra were fitted using a bi-exponential decay function:

$$y = y_o + A_1 \exp\left(-\frac{x - x_o}{\tau_1}\right) + A_2 \exp\left(-\frac{x - x_o}{\tau_2}\right)$$

$\tau_1$ and $\tau_2$ are the faster and slower time constants, respectively, and $A_1$, $A_2$ are the corresponding amplitudes. The TR-PL spectra taken at the both temperature range were fitted after 3 ps from the rise time to exclude the system response and any hot-carriers related effect.

**TR-PL spectra of the bound excitonic emission:**

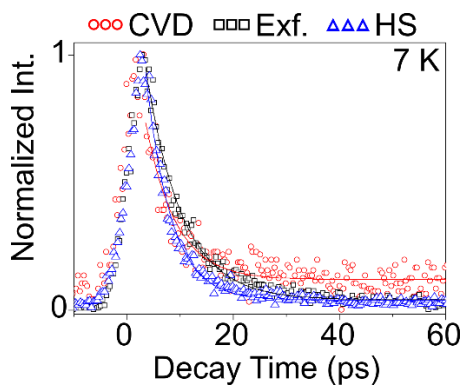

*Figure S7: Time resolved PL (TR-PL) spectra of the bound exciton ($X^-$) emission from the three regions measured at 7 K. The spectra were fitted using a single exponential decay function. The time constants from the CVD, Exf. and HS areas are 5.659 ± 0.369 ps, 6.801 ± 0.147 ps and 4.733 ± 0.096 ps, respectively. The hollow symbols represent the real signal and the solid lines are the fitted data.*

**Theoretical Calculation:**

**Band structure calculation**: DFT calculations is performed using the Materials Studio CASTEP (CAmbridge Serial Total Energy Package) version 2024, *ab initio* Total Energy Program (first principles methods using CASTEP).[1] We did not develop or validated any DFT codes employed in the calculation. Plane wave basis set cut-off in CASTEP as well as other parameters except K-point mesh were set to ultra-fine default option set in the software package/CASTEP module. Prior to the band structure calculation, we performed the geometry optimization (GO) constraining the lattice parameter obtained from experimental values. To obtain band structure of ML TMD, crystal was cleaved parallel to the layer (c* terminated) and then a vacuum slab > 20 Å was added along the c* to make the 1L TMD structures. In-plane lattice parameter of CVD MoSe$_2$ was fixed at $a$ = 3.25 Å and $a$ = 3.33 Å was considered for the Exf. MoSe$_2$. In DFT calculation, Atomic solver, Dirac (Author Chris J. Pickard, Cambridge University) in CASTEP module is used when requesting spin-orbit coupling and on the fly generated pseudopotentials. Self-consistent calculation performed for the test configuration for Se: $1s^2\ 2s^2\ 2p^6\ 3s^2\ 3p^6\ 3d^{10}\ 4s^2\ 4p^4$ and for Mo: $1s^2\ 2s^2\ 2p^6\ 3s^2\ 3p^6\ 3d^{10}\ 4s^2\ 4p^6\ 4d^5\ 5s^1$. Pseudo atomic calculation performed for Se $3d^{10}\ 4s^2\ 4p4$ and for Mo $4s^2\ 4p^6\ 4d^5\ 5s^1$.

Band structure calculation was performed considering the fine k-spacing in single point energy calculation corresponding to 20x20x1 supercell and spectral k-spacing of ~0.006Å$^{-1}$ along M-K-Γ-M. 2D band structure (E vs K$_x$, K$_y$) is obtained from DOS.band output file (eigen values vs K$_x$, K$_y$) in which k-point mesh in the computation is set corresponding to 100x100x1 supercell with unique 5000 k-points within k-space defined by the reciprocal space a*/2 (0 to a*/2) and b* (-b*/2 to b*/2). For GO and computing the ground state band structure of 1Ls MoSe$_2$, DFT-D (GGA+dispersion correction) method: Perdew-Bruke-Ernzerhof (PBE) GGA functional[2] is employed along with the dispersion correction (van der Waals correction accounted employing the dispersion correction for DFT) by Tkatchenko-Scheffler (TS) method.[3] After computation of the electronic band structure in CASTEP, scissors have applied to the band structure plot to match with the bandgap obtained from the RC spectroscopy measurements.

The WF calculations are performed based on the DFT by using the plane-wave (PW) method as implemented in Q$_{\text{UANTUM}}$ ESPRESSO code.[4-6] We use full-relativistic norm-conserving pseudopotentials for electron-ion interaction and PWs with a kinetic energy cutoff of 70 Ry to expand electronic wavefunctions. The interlayer vdW interaction is described by using the DFT-D3 dispersion correction method. The cutoff energy of the PW expansion is optimized to 70 Ry, which ensures the convergence of the system, and a 25x25x1 Γ-centered Monkhorst-Pack k-mesh grid is used in the BZ for structural optimization and performance calculations. The convergence thresholds for energy and atomic forces are fixed at $10^{-4}$ eV and $10^{-5}$ eV/Å, respectively. 15 Å vacuum layer along z-axis is added to interrupt the artifacts of the periodic boundary conditions, spin-orbit coupling (SOC) is considered in this work.

**Work function calculation**: The estimated work function (WF) is defined as $W^{CVD/Exf} = E_{Vacuum}^{CVD/Exf} - E_{F}^{CVD/Exf}$, where $W^{CVD/Exf}$, $E_{F}^{CVD/Exf}$ and $E_{Vacuum}^{CVD/Exf}$ denote the WF, Fermi level and vacuum level of CVD/Exf, respectively. The values of $W^{CVD/Exf}$ are estimated in Figure S8, which displays the average electrostatic potential in vacuum for both monolayers (CVD and Exf.). The estimation can be acquired via averaging the total electrostatic potential along the direction out-of-plane. Indeed, From the DFT calculation, the planar-averaged electrostatic potential is given as:

$$V(z) = \frac{1}{A_{xy}} \iint V(x,y,z) dxdy + E_0$$

where $A_{xy}$ is the transverse area of the MoSe$_2$ slab. The constant $E_0$ is an energy shift applied to make $V(z) = 0$ in the vacuum region. The WF of CVD and Exf. 1L MoSe$_2$ are 4.48 eV and 4.60 eV, respectively, with CVD exhibiting a 120 meV lower WF compared to Exf. Indeed, higher WF suggests that the electrons need high energy to get detached from the surface while smaller WF suggests that electrons can be easily removed from the surface. This difference implies that electrons on the surface of CVD 1L MoSe$_2$ require less energy to detach compared to those on Exf ML- MoSe$_2$. This WF values are in good agreement with previous theoretical calculation.[7]

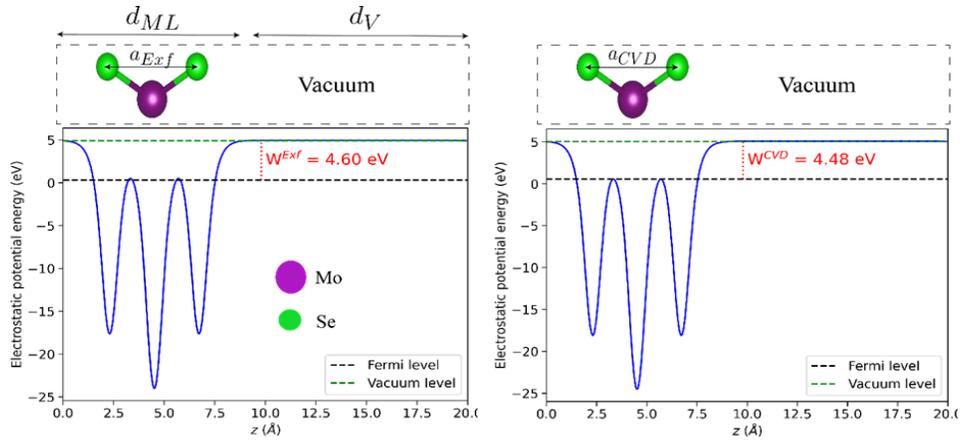

*Figure S8: Top panel are the side view of the atomic structure of MoSe$_2$ with lattice parameter $a_{Exf}$ in left and $a_{CVD}$ in right. $d_{ML}$ and $d_V$ are the monolayer and vacuum thickness. The bottom panel are the planar average of the electrostatic potentials (V(z)) distribution of the 1L MoSe$_2$. The values of $W^{CVD/Exf.}$ are indicated. The violet and green balls represent Mo, and Se atoms, respectively.*

**Reference**:

[1] Clark, S. J.; Segall, M. D.; Pickard, C. J.; Hasnip, P. J.; Probert, M. I. J.; Refson, K.; Payne, M. C. First Principles Methods Using CASTEP. 2005, 220 (5–6), 567–570.

[2] Perdew, J. P.; Burke, K.; Ernzerhof, M. Generalized Gradient Approximation Made Simple. Phys. Rev. Lett. 1996, 77 (18), 3865–3868.

[3] Tkatchenko, A.; Scheffler, M. Accurate Molecular Van Der Waals Interactions from Ground-State Electron Density and Free-Atom Reference Data. Phys. Rev. Lett. 2009, 102 (7), 073005.

[4] Giannozzi, P.; Baroni, S.; Bonini, N.; Calandra, M.; Car, R.; Cavazzoni, C.; Ceresoli, D.; Chiarotti, G.L.; Cococcioni, M.; Dabo, I. QUANTUM ESPRESSO: a modular and open-source software project for quantum simulations of materials. J. Phys.: Condens. Matter 21 395502.

[5] Giannozzi, P.; Andreussi, O.; *et al.* Advanced capabilities for materials modelling with Quantum ESPRESSO. J. Phys.: Condens. Matter 29 465901.

[6] Giannozzi, P.; Baseggio, O.; Bonfà, P.; Brunato, D.; Car, R.; Carnimeo, I.; Cavazzoni, C.; de Gironcoli, S.; Delugas, P.; Ferrari Ruffino, F.; Ferretti, A.; Marzari, N.; Timrov, I.; Urru, A.; Baroni, S. Quantum ESPRESSO toward the Exascale. J. Chem. Phys. 2020, 152 (15), 154105.

[7] Han-gyu, K.; Choi, H. J. Thickness dependence of work function, ionization energy, and electron affinity of Mo and W dichalcogenides from DFT and GW calculations. Phys. Rev. B 103, 085404.